# Network Subgraphs of heterogeneous Chinese credit system

Yingli WANG[a,b], Qingpeng ZHANG[c,*], and Xiaoguang YANG[a,b,*]

[a] MADIS, Academy of Mathematics and Systems Science, CAS

[b] University of Chinese Academy of Sciences

[c] School of Data Science, City University of Hong Kong

yinglwang5-c@my.cityu.edu.hk

qingpeng.zhang@cityu.edu.hk

xgyang@iss.ac.cn

**Abstract:**

In this paper, we analyze the subgraphs of the heterogenous Chinese guarantee network with five years' worth of real-world data from January 2007 to March 2012. We identify the motifs in 2- and 3-node subgraphs, and find that firms have special preferences to form mutual, triangle and 2-star guarantee relationships. Specifically, we consider the heterogeneity of firms and determine that firms with relatively large assets tend to form the sub-pattern of 2-out-stars and that small firms tend to form mutual guarantee relationships. Our results give the policy makers an in-depth understanding of Chinese guarantee firms' preferential behavior from the perspective of network subgraphs, and provide structural design instructions for future policy making to reduce the potential default and contagion risk of guarantee network.

**Keywords:** Heterogenous Complex Network; Network Motif; Functional Subgraph; Default Risk; Risk Contagion

## 1.Introduction

A financial network represents a collection of entities (i.e., firms, banks, trades or financial exchanges), linked by mutually beneficial business relationships [1, 2]. In particular, the guarantee relationship between two firms representing the responsibility of a firm (the guarantor) to assume the debt obligation of a borrower firm if that firm failed to meet its legal obligation of a loan (default). Through such interdependencies, firms are connected as a guarantee network. Loan guarantees could enhance the financing ability of firms, especially small and medium enterprises/firms (SMEs), thus facilitating the rapid growth in the economic upturn period.

* The research is supported by National Science Foundation of China Grants 71532013, and 71672163.
* Declarations of interest: none.
∗ Author contributions: Y.W., Q.Z., and X.Y. designed research; Y.W., and Q.Z. analyzed data and performed research; and Y.W., Q.Z., and X.Y. wrote the paper.
* The authors declare no conflict of interest.
* Y.W., Q.Z., and X.Y. contributed equally to this work.
* To whom correspondence should be addressed. E-mail: xgyang@iss.ac.cn; qingpeng.zhang@cityu.edu.hk

In this article, we discuss temporal and heterogenous Chinese guarantee networks constructed from guarantee relationship between firms. We focus more closely on the basic local structures of the guarantee networks, in particular, dynamic frequency of occurrence of 15 different subgraphs (2 different 2-node subgraphs and 13 different 3-node subgraphs) within the networks. In addition, we study the financial characteristics of these 15 different patterns, finding large firms are likely to be the guarantor in a 2-star pattern, and small firms probably get loan through mutual guarantee.

Furthermore, to our best knowledge, this was the first work using analysis of subgraphs to address the guarantee network default risk and contagion risk issues, finding the optimal local structures in the guarantee networks were 2-star and single-link patterns. Our local subgraph analysis of dynamic heterogenous Chinses loan guarantee networks was a useful approach for risk management of Chinese guarantee system.

Our empirical findings would facilitate the study of subgraph analysis of real-world heterogenous financial system and also enable better strategies to be developed for policy makers to identify firms with high default-risk and contagion ability.

## 2. Related work

Many of the complex network share same global statistical features, such as scale-free degree distribution [3], small-world property [4] and high clustering coefficient [40, 60]. The detailed explanations of these global properties are followed. If the degree distribution follows a power-law in the network, $p(k) \sim k^{-\lambda}$ [5]. Then we named it as a scale-free network, which indicates that nodes are only connected to a few edges, while there exist a few hub nodes that are densely connected with other nodes. In addition, many real-world networks usually exhibit a relatively small average shortest path length, indicating the small-world property [6-8]. As for the high clustering coefficient, which measures the extent to which a node's neighbors are also adjacent to each other. In real-world networks, nodes are likely to form such triads, resulting a higher average clustering coefficient of the networks. In social networks, it refers to the tendency that "friend of a friend is also a friend" [9, 10].

Notably, networks with the similar global characterizes may have significantly different local structures or networks with wildly different global features may demonstrate similar basic structures [11]. In reality, many researches have revealed there may be some relationships between global topological properties (small-world character, high clustering coefficient and scale-free feature) and key local subgraphs [12]. Therefore, in order to obtain a comprehensive understanding of a particular network, the local property is as important as global features.

In contrast to global topologies, the relative frequency of subgraphs is the statistic of local structural property, which can be calculated by the network motif detection. Motifs are recurrent small connected subgraphs or patterns, which displayed significantly higher frequencies in a particular network than would be expected for a randomized network [11, 13]. Recently this approach has been wildly used to uncover basic building blocks of networks. For example, motifs have already been found in networks from biochemistry, neurobiology, ecology, engineering food webs, neural networks and the World Wide Web [14-18].

### 3. Data and subgraph

In this research, we acquire a comprehensive dataset of loan guarantee records from a Chinese financial institution. The proprietary loan guarantee dataset is collected from the 19 major banks of China, that is the largest five state-owned commercial banks, the twelve joint equity commercial banks, and the two main policy banks (i.e., Import-Export Banks of China and China Development Bank).

Particularly, our dataset covers two important events in the global financial system: the global financial crisis of 2007-2008, and the implementation of the Chinese economic stimulus program from November 2008 to December 2010. As two rare natural experiments, these events offer a good chance to investigate the change of guarantee network under extreme exogenous shocks. from perspective of local subgraphs

The combinations of subgraphs with different structural properties give rise to the complexity of a network. To unveil combination patterns in guarantee networks, 2- and 3-node directed subgraphs are employed, which are described in the following.

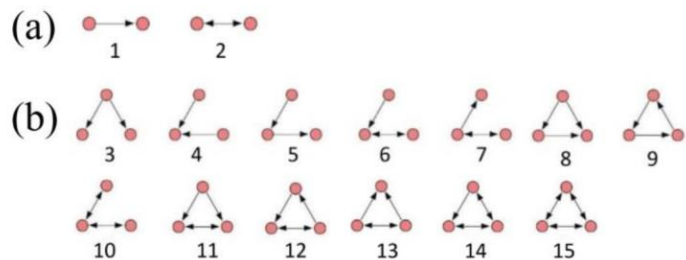

Fig. 2 All subgraphs of size 2 and 3 in directed networks.

In Figure 2, we demonstrate guarantee relationships represented by directed edges between 2 and 3 nodes used for our study. In total, there are 2 types of 2-node subgraphs and 13 types of 3-node subgraphs.

### 4. Results and Discussion of Motif Detection of Chinese Guarantee Network

In this section, we examine the static motif detection of the guarantee network. We analyze the motifs of guarantee network formed in each month, and summarize for the whole period. The motifs of the guarantee network are presented in Table 1. Some explanations of indexes in Table 1 are shown below, and the specific subgraphs of motifs can be found in Figure 2.

1) Frequency [Original] denoted the frequency of a particular subgraph in real guarantee network.
2) Mean-Freq [Random] was the mean frequency of a particular subgraph in randomized networks.
3) SD [Random] represented the standard deviation from the mean frequency in randomized networks.
4) Z-Score was the frequency in real network minus the random frequency divided by the standard deviation, so the larger Z-Score, the more statistically significant was the subgraph.
5) p-Value ranged from 0 to 1. The smaller the p-Value, the more significant was the motif.

**Table 1 Average motif detection of 63 guarantee networks.**

| Type | Frequency [original] | Mean-Freq [random] | SD [random] | Z-score | p-value |
| --- | --- | --- | --- | --- | --- |
| 1 | 0.856252 | 0.999979 | 3.05E-05 | -5041.85 | 1 |
| 2 | 0.143748 | 2.1E-05 | 3.05E-05 | 5044.148 | 0 |
| 3 | 0.470967 | 0.461571 | 6.27E-06 | 1530.45 | 0 |
| 4 | 0.18356 | 0.192128 | 8.08E-06 | -1134.72 | 1 |
| 5 | 0.162021 | 0.173016 | 9.76E-06 | -1197.04 | 1 |
| 6 | 0.058856 | 0.068375 | 1.08E-05 | -957.571 | 1 |
| 7 | 0.076654 | 0.086416 | 8.14E-06 | -1241.67 | 1 |
| 8 | 0.016996 | 1.71E-05 | 1.15E-05 | 1606.41 | 0 |
| 9 | 0.001107 | 4.49E-07 | 1.87E-06 | 614.2838 | 0 |
| 10 | 0.01399 | 0.018469 | 3.52E-06 | -1328.6 | 1 |
| 11 | 0.004124 | 3.54E-06 | 5.21E-06 | 865.9048 | 0 |
| 12 | 0.002691 | 1.51E-06 | 3.44E-06 | 816.2811 | 0 |
| 13 | 0.004884 | 1.72E-06 | 3.66E-06 | 1372.92 | 0 |
| 14 | 0.00334 | 1.66E-06 | 3.59E-06 | 968.8621 | 0 |
| 15 | 0.000811 | 3.69E-09 | 9.51E-08 | 3186.1 | 0 |

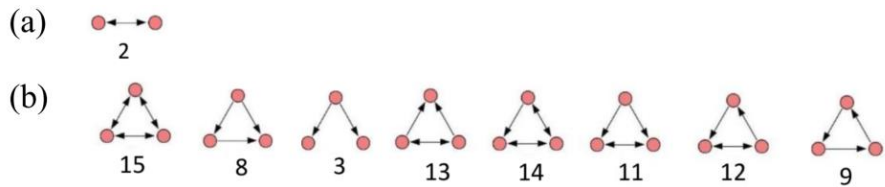

**Fig. 3 Ranking of motif based on their Z-score (from high to low).**

Firstly, there are nine motifs (type 2, 3, 8, 9, 11, 12, 13, 14 and 15) in the guarantee network. According to the approach of motif detection, the larger Z-score of a particular motif represents the more prevalence of such subgraph in real network. Thereby, in Figure 2, we sort the nine motifs by Z-score from highest to lowest.

Accordingly, Figure 2 (a) and (b) showed the motif in 2- and 3-node subgraphs, respectively, in which the motifs are sorted from highest to lowest by Z-score. Therefore, from Figure 2, we can see the favorite local guarantee patterns of 2 and 3 firms.

Secondly, as we all know, network motifs are patterns of interconnections significantly recurring among firms in guarantee networks than in randomized networks, that is to say, the motifs in guarantee network could represent firms' preferential behavior or favorite patterns when forming guarantee relationship. Concretely, the motif of type 2 indicates the mutual guarantee is prevalent among firms. Motif of type 2 represents the popularity of two-star guarantee pattern. The recurrence of triangle guarantee relation can be discovered by motifs of type 8, 9, 11, 12, 13, 14, and 15.

## 5. Conclusions

To our knowledge, we were the first one to analysis the heterogenous and dynamic guarantee networks from the perspective of local structure. Firstly, we found the statistically important and functionally important subgraphs in the Chinese guarantee networks. Firms in the guarantee network have a special preference to form local mutual and triangle relationships. Additionally, the network motifs have been widely used in the biochemistry, neurobiology, ecology, and engineering system [14-16] [17, 18], which are considered as homogeneous networks with equivalent objectives. Conversely, the firms in our guarantee networks are heterogeneous. Therefore, we made a further analysis of the correlation between firms' financial attributes and favorable subgraphs. We find large firms tend to be guarantor in 2-star motif, and small firms are likely to form mutual guarantee relationship to acquire loan from banks. In addition, we demonstrate an approach how to fast locate the high default firms in the guarantee network. At last, we find the subgraph of single-link and 2-star are with low contagion ability. Thereby, our study may shed light on the preferential guarantee patterns among firms, and optimal structure design to reduce default or contagion in the guarantee networks. Our analysis could help the government and bank to monitor default spread status and provided insight for taking precautionary measures to prevent and solve systemic financial risk.